\documentclass{eas}
\usepackage{graphicx}
\begin{document}

\title{Chemodynamics of Lyman alpha emitters, Lyman break galaxies and elliptical galaxies} 
\author{Masao Mori}\address{Institute of Natural Sciences, Senshu University, Kawasaki, 
Kanagawa 214-8580, Japan}
\author{Masayuki Umemura}\address{Center for Computational Sciences, University of Tsukuba, 
Tsukuba, Ibaraki 305-8577, Japan}
\begin{abstract}
We report an ultra-high-resolution simulation that follows evolution from the earliest 
stages of galaxy formation through the period of dynamical relaxation. The bubble 
structures of gas revealed in our simulation ($< 3\times 10^8$ years) resemble closely 
the high-redshift Lyman $\alpha$ emitters (LAEs). 
After $10^9$ years these bodies are dominated by stellar continuum radiation and look like 
the Lyman break galaxies (LBGs) known as the high-redshift star-forming galaxies at which 
point the abundance of elements appears to be solar. 
After $1.3\times10^{10}$ years, these galaxies resemble present-day ellipticals. 
The comparisons of simulation results with the observations of elliptical galaxies 
allow us to conclude that LAEs and LBGs are infants of elliptical galaxies or bulge 
systems in the nearby universe.
\end{abstract}
\maketitle
\section{Introduction}
Recent progress in observational devices and techniques has enhanced our knowledge of 
formation and evolution of galaxies on a firm statistical basis. Optical observations 
have revealed the presence of a number of LAEs at redshifts of $z \geq$ 3 
(Dey {\em et al.} 1998; Taniguchi {\em et al.} 2003 and references therein) 
as well as LBGs at redshifts of $3 \leq z \leq 6$ (Steidel {\em et al.} 1996; 
Giavalisco 2002 and references therein). 
A good fraction of LBGs exhibit Ly$\alpha$ emission lines strong enough to be detected 
by narrow band observations (Shapley {\em et al.} 2003). 
Since LAEs and LBGs are quite young, they could hold direct interpretable information on 
the early chemical enrichment of galaxies, contrary to present-day galaxies which have 
undergone intense interstellar medium (ISM) recycling, thus erasing most of the early 
chemical history.

The modern paradigm of the galaxy formation, based on the cold dark matter hypothesis, 
deduces that galaxies formed hierarchically in a bottom-up fashion, where a larger system 
results from the assembly of smaller dark matter halos. Baryonic gas falls into the 
gravitational potential of dark matter halos, and condenses rapidly as a result of the 
radiative cooling for atoms or molecules. The dense cooled gas clouds are the birth sites 
of stars, and massive stars born there explode as type II supernovae (SNe) in a few times 
$10^7$ years. 
Continual multiple SNe generate hot bubbles enriched with heavy elements and shocked dense 
shells. Hence, to study the chemical enrichment, it is crucial to resolve accurately the 
thermalization of the kinetic energy released by multiple SNe and the mixing of heavy 
elements.
So far, the theoretical models of galactic chemical evolution have often assumed the 
homogeneous ISM (one-zone model), with the instantaneous and perfect mixing of heavy 
elements synthesized in SNe. However, the energy input and metal ejection by SNe are 
likely to proceed in an inhomogeneous fashion (Mori {\em et al.} 2004).
Thus, simulations that can resolve SN remnants are required to properly model the 
chemical evolution of primordial galaxies. In this paper, we perform ultra-high 
resolution hydrodynamic simulations of a very large burst of multiple SN explosions in 
a forming galaxy. Also, by incorporating spectro-photometric modeling with the simulations, 
the results can be directly compared to the observations on high-redshift forming 
galaxies. The outline of this paper is as follows. In Section 2, the numerical method 
is described. In Section 3, we present the simulation results and in Section 4 the 
possible link among LAEs, LBGs and elliptical galaxies is discussed.
The full description of this study can be found in Mori \& Umemura (2006).

\section{Numerical method}
Our simulation uses a hybrid $N$-body/hydrodynamics code which is applicable to 
a complex system consisting of dark matter, stars and gas. The gas is allowed 
to form stars and is subject to physical processes such as the radiative cooling
and the energy feedback from SNe. The gas is assumed to be optically thin and in 
collisional-ionization equilibrium. Radiative cooling is included self-consistently
with metallicity, using the metallicity-dependent cooling curves by Sutherland \&
Dopita (1993).
The collisionless dynamics for dark matter particles and stars is treated by the 
$N$-body method and the gas dynamics is pursued by a three-dimensional AUSM-DV 
scheme that can treat shocks with high accuracy (Mori {\em et al.} 2002). 
Since this scheme has a great advantage due to the reduction of numerical 
viscosity, fluid interfaces are sharply preserved and small-scale features can be 
resolved.

Stars are assumed to form in rapidly cooling and Jeans unstable regions at a rate 
which is inversely proportional to the local dynamical time (see Mori {\em et al.} 
1997, 1999). When a star particle is formed, we identify this with 
approximately $10^4$ single stars and distribute the associated mass of the star 
particle over the single stars according to Salpeter's initial mass function.
The lower and upper mass limits are taken as 0.1 and 100 $M_\odot$, 
respectively. When a star particle is formed and identified with a stellar assemblage 
as described above, stars more massive than $ 8 ~M_\odot$ start to explode as Type II SNe 
with the explosion energy of $10^{51}$ ergs and their outer layers are blown out with 
synthesized metals leaving the remnant of $1.4 ~M_\odot$. Consequently, once a new star 
particle is formed, the energy, metals and material from Type II SNe are subsequently 
supplied to 8 cells surrounding SN region. We compute the chemical evolution using 
the calculations of stellar nucleosynthesis products by Tsujimoto {\em et al.} (1995). 
A mass of $2.4 ~M_\odot$ of oxygen is ejected from a Type II SN explosion.

\section{Simulation}

Following a standard cold dark matter cosmology with cosmological constant ($\Lambda$CDM), 
where we assume $\Omega_M = 0.3, \Omega_\Lambda = 0.7, \Omega_b = 0.04$, and a Hubble 
constant of $H_0 = 70$ km s$^{-1}$ Mpc$^{-1}$, we consider the dynamical and chemical 
evolution of a protogalaxy with the total mass of $10^{11} M_\odot$. We assumed that the 
total mass of gaseous matter is $1.3\times10^{10} ~M_\odot$ initially. If we suppose a 
$2\sigma$ density fluctuation, it decouples from the cosmic expansion and begins to contract 
at redshift $z = 7.8$ with the radius of 53.7 kpc. 
The angular momentum is provided by a uniform rotation characterized by a spin parameter 
of $\lambda=0.05$. Prior to this galaxy-scale fluctuation, 
subgalactic dark halos collapse and are virialized. According to the $\Lambda$CDM cosmology, 
twenty subgalactic condensations with mass of $5.0 \times 10^9 M_\odot$ and radius of 
8.6 kpc are distributed within the galaxy-scale fluctuation. The subgalactic virialized 
halos are assumed to follow the Navarro-Frenk-White density profile (Navarro {\em et al.} 1997). 
The hydrodynamic processes are pursued with $1024^3$ grid points. The simulation box has a 
physical size of 134 kpc and the spatial resolution is 0.131 kpc.

Fig. 1 shows the result for the time sequence of the chemical enrichment, where the 
distributions of the logarithmic density, the temperature and the velocity, and oxygen 
abundance [O/H], are presented until $10^9$ years. In the first $10^8$ years, stars are 
formed in high-density peaks within subgalactic condensations and the burst of star 
formation starts. Then, massive stars in the star forming regions explode as SNe one 
after another, producing expanding hot bubbles surrounded by cooled dense shells. The gas 
in the vicinity of SNe is quickly enriched with ejected heavy elements, but a large amount 
of gas still retains low heavy-element abundance. Consequently, the metallicity distribution 
becomes highly inhomogeneous, where gas enriched as $-5 \leq [O/H] \leq -1$ coexists with 
virtually primordial gas. Since the density of the ISM is lower in the outer regions of 
subgalactic condensations, the expansion of hot bubbles is accelerated there and SN-driven 
shocks collide with each other to generate super-bubbles of $\sim 50$ kpc, and the surrounding 
high-density, cooled ($T = 10^{4}$ K) shells form at $3\times10^8$ years. The hot bubbles 
expand further by subsequent SN explosions, and the shells sweep up the partially 
enriched ambient gas. The gas density in dense shells increases owing to the efficient 
radiative cooling mainly through collisional excitation of neutral hydrogen. 

After $5\times10^8$ years, the hot bubbles blow out into the intergalactic space. 
This SN-driven outflow is an efficient mechanism to enrich the intergalactic 
medium with heavy elements over a large cosmological volume (see Mori {\em et al.} 2002). 
On the other hand, the dense shells undergoes hydrodynamic instabilities induced by 
shell-shell interactions and radiative cooling, eventually fragmenting into cold 
filaments and blobs. These interactions are giving rise to an intricate multiphase 
structure in the inner halo, where $10^{6-7}$ K gas coexists with a cooler $10^4$ K phase 
from which it is separated by cooling interfaces. New stars are born in this enriched gas 
and again heavy elements are ejected from subsequent SNe. The rightmost panels show the 
structure at $10^9$ years. By this stage, the ISM is recycled repeatedly and about 72\% of 
the initial gas has been processed into stars. Eventually, some amounts of cool, dense 
filaments are left at the center. But, the most of volume is filled with rarefied gas 
($n \leq 10^{-4}$ cm$^{-3}$) that has intermediate temperature 
($10^{4.5}$ K $\leq T \leq 10^{6.5}$ K). At this epoch, the mixing of heavy elements is 
nearly completed.

\begin{figure}[t]
 \begin{center}
  \includegraphics[width=120mm, height=90mm]{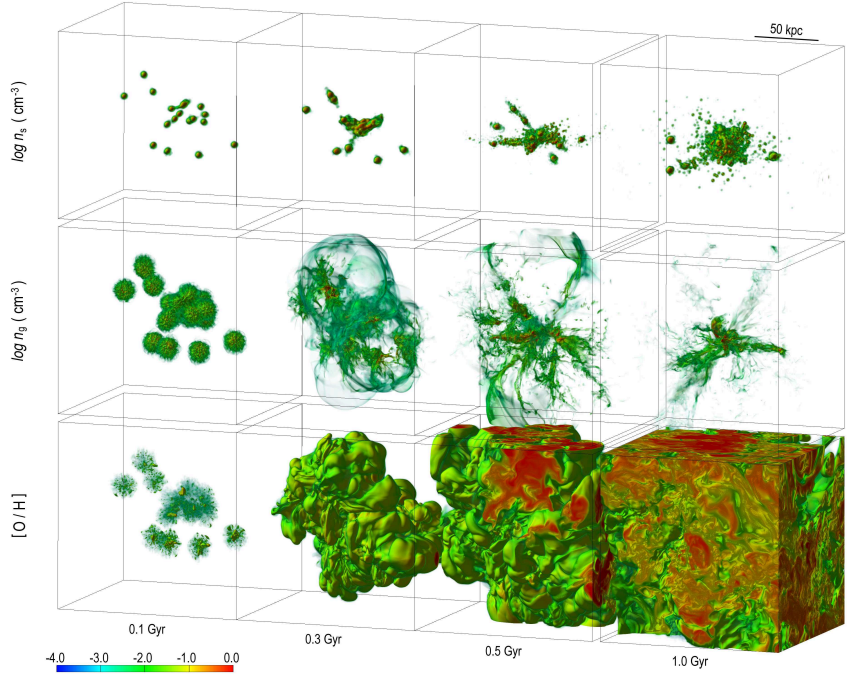}
  \caption{Spatial distributions of the stellar density, the gas density, and the oxygen 
abundance [O/H] of gas. The four panels in each column depict the time evolution of the 
simulation results until 1 Gyr.}
 \end{center}
\end{figure}

\section{Discussion and conclusion}

\begin{figure}[t]
 \begin{center}
  \includegraphics[width=100mm]{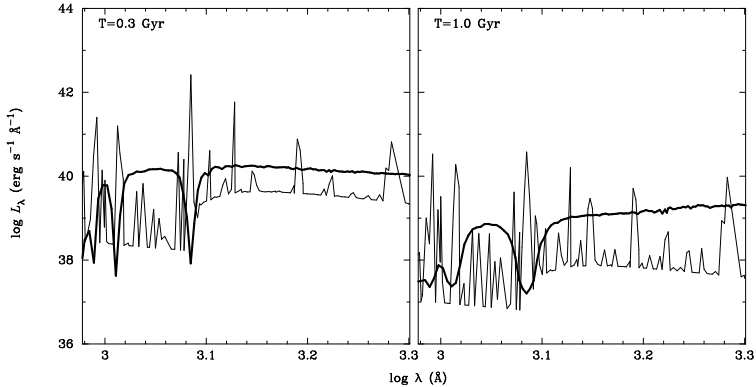}
  \caption{Predicted emission properties of the gas components assuming for an optically 
thin gas in collisional-ionization equilibrium (thin lines), and that of the stellar 
components using the evolutionary stellar population synthesis (thick lines).}
 \end{center}
\end{figure}

The emission properties of the gas components are calculated for an optically-thin, 
collisionally-ionized gas using MAPPINGIII code by Sutherland \& Dopita (1993), and 
those of the stellar components are calculated using the evolutionary stellar population 
synthesis code PEGASE by Fioc \& Rocca-Volmerange (1997).
In practice, to obtain the SED, we sum up the SED of each grid point for the gas 
components and each star particle for the stellar components. 
Thus, the present simulations can be directly compared to the observations.

Fig. 2 shows the spectral energy distribution (SED) of the simulated galaxy. 
Thin lines and thick lines depict, respectively, the gas emission and the stellar emission. 
The Ly$\alpha$ emission comes mainly from high-density cooling shells, and its luminosity 
is more than $10^{43}$ erg s$^{-1}$ in the first $3\times10^8$ years. This Ly$\alpha$ 
luminosity perfectly matches that of observed LAEs (Matsuda {\em et al.} 2004). 
Furthermore, we find that the physical extent of $\sim 100$ kpc 
and the bubbly structure produced by multiple SNe are quite similar to the observed features 
in Ly$\alpha$ surface brightness distribution of LAEs. 
After $3\times10^8$ years, the Ly$\alpha$ luminosity quickly declines to several 
$10^{41}$ erg s$^{-1}$ that is lower than the observed level. This result suggests that 
LAEs can correspond to an early SN-dominated phase before $3\times10^8$ years. 

As seen in Fig. 2, after the sparkling phase of a primeval galaxy, the SED is dominated 
by stellar continuum emission, since the emission from cooling gas decreases immediately 
owing to the leak of explosion energy through the blowout of super-bubbles. The galaxy in 
this phase is featured with diffuse, asymmetric structures, and outflows of $100\sim500$ 
km s$^{-1}$. The total mass of long-lived stars is $9.3\times10^9 M_\odot$, and the mass of 
$1.5\times10^9 M_\odot$  is involved in the outflows at $z=3$. These features look quite 
similar to those observed for LBGs (Adelberger {\em et al.} 2003). 
In the light of such properties, the simulated post-starburst galaxy that has the age of 
$10^9$ years can correspond to LBGs.

The following dynamical evolution is studied by an $N$-body simulation with one million 
particles, to explore the end-product of this galaxy. As a result, it is found that the 
assembly of sub-condensations and the virialization of the total system are almost completed 
in $3\times10^9$ years, so that the system becomes in a quasi-equilibrium state. The 
resultant stellar system forms a spherical virialized system. 
The projected surface brightness distributions have a large central concentration that 
well accords with de Vaucouleurs' $r^{1/4}$ profile and the resultant absolute magnitude in 
blue band and visual band are $M_B=-17.2$ mag and $M_V=-18.0$ mag, respectively. 
The colour $U-V=1.15$ and $V-K=2.85$ are consistent with the colour-magnitude relation of 
elliptical galaxies in Coma cluster of galaxies (Bower {\em et al.} 1992). 
Furthermore, the combination of the surface brightness, the effective radius $r_e=3.97$ kpc, 
and the central velocity dispersion $\sigma_0=133$ km s$^{-1}$ is on the fundamental plane 
of elliptical galaxies within their scatters (Djorgovski \& Davis 1987).
These comparisons of the simulation results with the observations allow us to derive 
a significant conclusion that LAEs and LBGs are progenitors of present-day elliptical 
galaxies, and the on-going, major chemical enrichment phases.

\section*{Acknowledgements}

This work was supported in part by the Grant-in-Aid of the JSPS, 14740132, and by 
Grants-in-Aid of the MEXT, 16002003. The computations reported here were performed 
on the Earth Simulator at the JAMSTEC, the SPACE at Senshu University, and
the computational facilities at CCS in the University of Tsukuba.


\end{document}